\documentclass[oldversion]{aa}

\usepackage{amsmath}
\usepackage{graphicx}
\usepackage{txfonts}
\usepackage{natbib}
\usepackage{dsfont}
\usepackage{color}

\definecolor{lightred}{rgb}{1.0,0.0,0.0}
\definecolor{lightgreen}{rgb}{0.0,1.0,0.0}
\definecolor{lightblue}{rgb}{0.25,0.25,1.0}
\definecolor{white}{rgb}{1.0,1.0,1.0}

\newcommand{\e}{\mathrm{e}}
\newcommand{\ii}{\mathrm{i}}
\renewcommand{\d}{\mathrm{d}}

\bibpunct{(}{)}{;}{a}{}{,}

\begin{document}

\title{Arcfinder: An algorithm for the automatic detection of
  gravitational arcs}
\titlerunning{Arcfinder}
\author{Gregor Seidel \and Matthias Bartelmann}
\authorrunning{Gregor Seidel \& Matthias Bartelmann}
\institute{Zentrum f\"ur Astronomie, Universit\"at Heidelberg, ITA,
  Albert-Ueberle-Stra{\ss}e 2, 69117 Heidelberg, Germany}

\date{Received <date> / Accepted <date>}

\abstract
 {We present an efficient algorithm designed for and capable of
  detecting elongated, thin features such as lines and curves in
  astronomical images, and its application to the automatic detection
  of gravitational arcs. The algorithm is sufficiently robust to
  detect such features even if their surface brightness is near the
  pixel noise in the image, yet the amount of spurious detections is
  low. The algorithm subdivides the image into a grid of overlapping
  cells which are iteratively shifted towards a local centre of
  brightness in their immediate neighbourhood. It then computes the
  ellipticity for each cell, and combines cells with correlated
  ellipticities into objects. These are combined to graphs in a next
  step, which are then further processed to determine properties of
  the detected objects. We demonstrate the operation and the
  efficiency of the algorithm applying it to HST images of galaxy
  clusters known to contain gravitational arcs. The algorithm
  completes the analysis of an image with $3000\times3000$ pixels in
  about 4~seconds on an ordinary desktop PC. We discuss
  further applications, the  method's remaining problems and
  possible approaches to their solution.
}

\maketitle

\section{Introduction}

Gravitational arcs are an important diagnostic for the innermost mass
distribution in galaxy clusters, and thus they are an important
indirect diagnostic for a variety of questions in cosmology and
structure formation. So far, they have been detected almost
exclusively by visual inspection of cluster images, although
algorithms for their automated search have recently been proposed
(\citealt{franklenzen,assafhoresh,christophealard}).

There are several good reasons to search for ways to detect arcs in an
automated fashion. First, the unambiguous definition of arc samples
calls for an objective and reliable way to detect arcs and quantify
their properties. Second, wide-field surveys such as the
Canada-France-Hawaii Legacy Survey combine huge data fields with
sufficient depth to reach below the detection limit for arcs with
their typically low surface brightness. For arc statistics and its
potential importance for cosmological studies or investigations of
cluster dynamics, objectively searching for arcs in these wide-field
images promises an important step forward. However, scanning
wide-field data by eye covering hundreds of square degrees for arcs
with their typical widths of $\lesssim1''$ and lengths of
$\lesssim10''$ appears as a hopeless endeavour.

Automated arc searches can be conducted where clusters have previously
been identified, e.g.~through their optical appearance or X-ray
emission, but they can and should profitably be extended to blind
searches on large areas. Obviously, such goals can only be pursued if
an algorithm for automated arc detection is available.

The surface brightness of gravitational arcs is typically close to the
background, which may vary across the image. This is one reason why we
propose a new algorithm here rather than using one of those that were
described and implemented earlier. For instance, the anisotropic
diffusion underlying the algorithm by Lenzen et al. bears the risk of
creating elongated features from the noise which may then be hard to
distinguish from real arcs. Moreover, we aim at an algorithm which is
simple, thus presumably robust, and fast enough to be applied to large
data fields. This rules out better studied techniques for the
detection of line-like features in images, such as the Hough
transform. Finally, the algorithm must be capable of distinguishing
artifacts such as diffraction spikes or parts thereof from arcs. As we
shall show below, the algorithm proposed here does indeed satisfy
these criteria.

\section{Approach}

We begin the description of our algorithm by summarising what it is
supposed to achieve. Next, we shall explain the numerical methods
used and the reasons for employing them.

\subsection{General description}

We want the algorithm to identify features in astronomical images with
the following properties:

\begin{enumerate}

\item having a higher mean pixel intensity than the surrounding
  background, including faint features with a mean pixel intensity
  well within the domain of the pixel noise where segmentation by an
  intensity threshold is impractical;

\item being much longer than wide, more precisely being elongated in the
  sense that it is possible at each point within a feature to find
  a local direction along which the feature's intensity changes much
  less than perpendicular to it;

\item being extended in the sense that the feature's local curvature,
  i.e.~its change of orientation per unit length along its principal
  direction, is small.

\end{enumerate}

According to this definition, a ring with a circumference of 100 units
(e.g.~pixels) and a width of 5 units should be detected as one
feature, the equally narrow sides of an equilateral triangle of
comparable size should be detected as three features because of the
high curvature at its vertices, and an approximately circular object
with a length-to-width ratio near unity should not be detected at all.

\begin{figure}
  \includegraphics[width=\hsize]{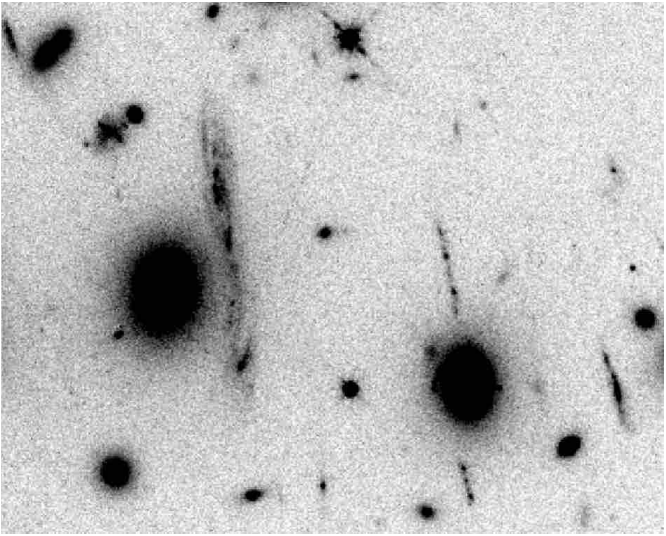}
\caption{Small section of an HST/WFPC2 observation of Abell~2390
  showing three arcs. Considerable substructure is visible in the
  straight arc on the left side.}
\label{image:a2390_substructure}
\end{figure}

Bearing strong gravitational lensing in mind, such features may be
considerably substructured, which argues for choosing a detection
scale limiting the size of detections to features exceeding a certain
length, while structures on smaller scales should be ignored. Another
reason for limiting the detection scale from below is the
contamination of images with numerous elongated features generated by
noise which are typically only a few pixels long.

A straightforward approach to remove small-scale structures is to 
initially convolve the image with a sufficiently large, isotropic 
kernel function, and to apply all further computations to the convolved 
image. However, while the convolution enhances the signal-to-noise 
ratio per pixel, it tends to blur and remove faint and thin features,
besides being computationally expensive. It is interesting to note, and
easy to see in a Fourier representation, that in an image convolved with
a Gaussian kernel, fluctuations of decreasing size become increasingly
unlikely.

We thus follow a different approach. We first directly compute the
particular image property we are interested in as local average within
areas corresponding to the detection scale. This avoids the blurring
effect of an initial Gaussian convolution, but retains its positive 
aspects, in particular the suppression of substructures and the
enhancement of the signal-to-noise ratio. It also significantly
decreases the execution time because averages are only calculated
for a limited set of pixels instead of all image pixels, assuming slow
variations of the image property spatially close to the pixels in the
set. Since they are interdependent, both the particular image property 
chosen for the algorithm and the selection of the set of pixels will 
be detailed later.

Since different local image properties, for example the
smoothed intensity or its gradient, will typically
vary on different scales, a smooth spatial variation on a single scale
cannot generally be assumed for all local image properties one may be
interested in. For example, while the intensity gradient will turn by
$180^\circ$ when the position is shifted by a few pixels to opposite
sides of a local intensity maximum, the smoothed intensity will hardly
be affected by the same spatial shift. The specific characteristics
chosen for detecting features must thus be taken into account in the
original selection of the pixel set.

In the following, pixels combined in a set, together with their
respective neighbourhoods used to compute local averages of certain
image properties, are called \emph{cells}. The cells' centre coordinates
vary and initialise the pixel set prior to each step of the algorithm.
Even when averaged within a single cell, local image properties turn out
to be insufficient for identifying faint features in presence of noise.
Thus, analogous to lines of magnetic flux visible in the pattern of iron
filings sprinkled on a slab of glass, the similarity of properties among
groups of cells in the presence of a feature is used as the main
criterion in our algorithm.

\subsection{Ellipticities and cell transport}

The average local pixel intensity of a feature can vary along its
length, with the only limitation that it must exceed the brightness of
the background. Thus, it is clearly not a good image property to use,
even if most of the background could be ignored after a preselection of
possible features above a minimal brightness threshold. Besides, this
is problematic in itself because valid features may be fainter than
typical variations in the image background.

However, since the local curvature in the direction of a valid feature 
is small by definition, and a flat background has no preferential 
orientation, directions in the local brightness pattern are a sound
criterion for a detection based on local correlations between image
pixels. Since the brightness gradient points into opposite directions
on opposite sides of the feature, spatially averaging over it will
eliminate the signatures of faint features. Using either the gradient
or the structure tensor, which is the Cartesian product of the gradient
with itself, is thus a problematic detection method for thin features
and reasonably large scale sizes.

Instead, one can compute the direction from the ellipticity given by the
second brightness moments. Let $I(\vec{x})$ be the intensity at the
position $\vec{x}=(x_1,x_2)$ and $q$ an appropriately chosen weight
function. Then, the weighted centre-of-brightness in an area $A$ is
\begin{equation}\label{eq:xbar}
  \bar{\vec{x}}=\frac{\int_{A}\vec{x}q(I(\vec{x}))\,\d^2x}
                     {\int_{A}q(I(\vec{x}))\,\d^2x}\;.
\end{equation}
For example, $q(I(\vec{x}))=I(\vec{x})$ will return the unweighted
centre-of-light. The tensor of second brightness moments has
the components
\begin{equation}\label{eq:Qij}
Q_{ij}=\frac{\int_{A}(x_i-\bar{x}_i)(x_j-\bar{x}_j)q(I(\vec{x}))\,\d^2x}
              {\int_{A}q(I(\vec{x}))\,\d^2x}
\end{equation}
with $i,j\in\{1,2\}$. The complex ellipticity is
\begin{equation}\label{eq:chi}
  \chi=\frac{Q_{11}-Q_{22}+2\ii Q_{12}}{Q_{11}+Q_{22}}\;.
\end{equation}

If $I(\vec{x})$ has elliptical isophotes in $A$ with an axis ratio of
$r\le1$, then $\chi=(1-r^2)/(1+r^2)\,\exp(2\ii\vartheta)$,
where $\vartheta$ is the orientation of the major axis of the
elliptical isophotes relative to the $x_1$ axis. $\chi$ is
invariant under rotations of $\pi$, as it should be because they
leave ellipses unchanged.

A vector $\vec e$ pointing into the direction of the major axis can be
obtained by bisecting the phase angle of $\chi$ (see
Appendix~\ref{ap:orientation}),
\begin{equation}\label{eq:e}
  \vec{d}=
  \begin{cases}
    (\chi_1+|\chi|,\chi_2) & \mbox{for}\quad\chi_1\ge0 \\
    (\chi_2,|\chi|-\chi_1) & \mbox{for}\quad\chi_1<0   \\
  \end{cases}\quad\mbox{and}\quad
  \vec{e}=\frac{\vec{d}}{|\vec{d}|}\;.
\end{equation}
The orientation obtained in this way is more stable against changes
in the scale size than gradient-based methods and has a higher
signal-to-noise ratio because the second brightness moments are
typically computed near a feature's centre-of-brightness.

For recognising multiple features in an astronomical image without
\emph{a priori} information on their positions, it is obviously
necessary to process the entire image initially. Our algorithm
starts with cells evenly distributed and convering the image
completely. Higher sensitivity can be achieved if neighbouring
cells overlap, but a large overlap should be avoided to reduce the
number of necessary operations on the image.

The first step in measuring the ellipticity is finding the centre of
brightness as in Eq.~(\ref{eq:xbar}). For a feature at or near the
edge of a cell, the centre cannot be accurately determined in a single
computation of $\bar{\vec{x}}$. For example, when applied once to
a smoothed vertical line with a Gaussian brightness profile on an
already estimated background $I_0$,
\begin{equation}
  I(\vec{x})=I(x_1)=I_0+\e^{-x_1^2/2\sigma^2}\;,
\end{equation}
with a weight function $q(I)=I-I_0$ and an area $A$ extending from $0$
to $4\sigma$ in the $x_1$ direction, Eq.~\ref{eq:xbar} yields
\begin{eqnarray}
  \bar{x}_1&=&\left(
    \int_0^{4\sigma}x_1\e^{-x_1^2/2\sigma^2}\d x_1
  \right)\left(
    \int_0^{4\sigma}\e^{-x_1^2/2\sigma^2}\d x_1
  \right)^{-1}\nonumber\\&=&-\left(
    \left.\sigma^2\e^{-x_1^2/2\sigma^2}\right|_0^{4\sigma}
  \right)\left(
    \left.\sigma\sqrt{\frac{\pi}{2}}
    \mathrm{erf}(\frac{x_1}{\sigma\sqrt{2}})\right|_0^{4\sigma}
  \right)^{-1}\approx0.798\sigma\;.
\end{eqnarray}
The result improves if $\bar{\vec{x}}$ is computed several times,
each time shifting the cell's centre to the position $\bar{\vec{x}}$
found in the last iteration. In our example, this yields approximate
centre positions $\bar{x}_{1,0}=2\sigma$,
$\bar{x}_{1,1}\approx0.798\sigma$,
$\bar{x}_{1,2}\approx0.210\sigma$,
$\bar{x}_{1,3}\approx0.048\sigma$, $\bar{x}_{1,4}\approx0.011\sigma$ and
$\bar{x}_{1,5}\approx0.002\sigma$ after five iterations, where the
second subscript denotes the iteration number.

\begin{figure}[!ht]
  \includegraphics[width=\hsize]{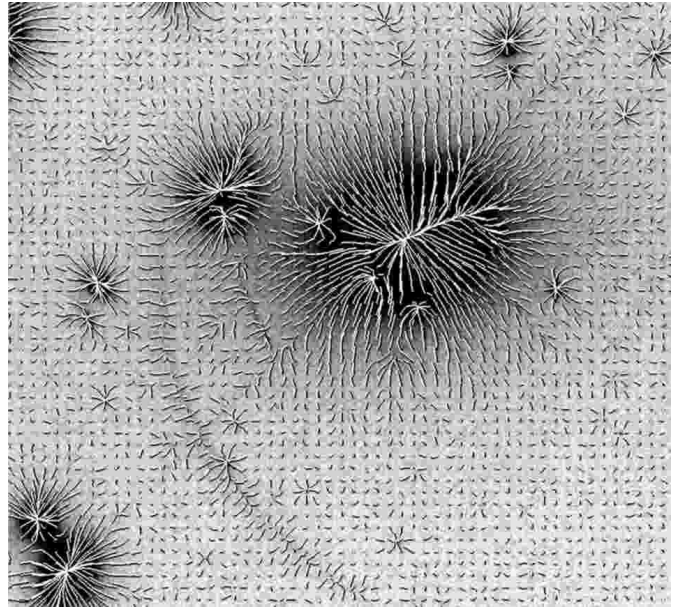}
\caption{Another section from the HST/WFPC2 image used for
  \figurename~\ref{image:a2390_substructure}, now centred on the optical
  emission of Abell~2390, showing cell paths leading to the closest
  local centres of brightness. The faint diagonal line crossing the 
  image from the bottom left to the top right is a diffraction spike
  from a nearby foreground star.}
\label{image:a2390_center:paths}
\end{figure}

In order to allow meaningful later correlation measurements between
cells, shifting initially neighbouring cells to exactly the same
final position must be avoided, since there the ellipticities would
then necessarily be equal regardless of a feature's real properties.
Ideally, the cells should instead be distributed equally along the
feature's main axis at the end of their iterative motion. More
precisely, they should end up on or near the feature's \emph{ridge
line}, i.e.~the line parallel to its local directions along the
brightness maxima measured perpendicular to these directions.

Carrying out too many iteration steps may create paths converging on
local maxima due to substructures in the feature, while too
few may leave $\bar{\vec{x}}$ far from the ridge line. In the worst
case, this may result in an orientation $\vec{e}$ perpendicular to the
feature for cells in areas characterised by a positive second
derivative in the intensity profile perpendicular to the ridge line
such as ``valleys'' between neighbouring maxima. This is because
the orientation $\vec{e}$ tends to be perpendicular to the direction
of maximum intensity curvature (the secondary brightness moments
$Q_{ij}$ vanish for an intensity linearly dependend on both
coordinates). In the case of a Gaussian intensity profile, the effect
would be visible if cells ended up more than one $\sigma$ away from the
ridge line. Thus, some compromise has to be achieved. Typically, about
three to four iterations have the desired effect.

It is important to note that this method is invariant under a linear
scaling of the relevant weighted intensities. In the current
implementation of the algorithm, $q(I)=\mathrm{max}(I-\bar{I},0)$ is
used for the determination of the centre-of-brightness, where
$\bar{I}$ is the mean intensity in $A$, leaving the method also
invariant under a constant offset to $I$ and, more importantly,
leading to a reasonably fast convergence towards the ridge line.

\subsection{Correlation of neighbouring cells}

Once the cells have been shifted to their new centre positions
$\vec{x}$, normalised directional vectors $\vec{e}$ can be readily
calculated using Eqs.~(\ref{eq:Qij}) through (\ref{eq:e}), where the
weight function $q(I)=I-\bar{I}$ is used for determining the
second brightness moments $Q_{ij}$. Cells not initially near a feature
and on a mostly flat background move randomly by short distances
due to the noise, and their orientations will also be random, while
cells originally located near a feature will arrive at different
positions along its ridge, with very similar local orientations.

\begin{figure}
  \includegraphics[width=\hsize]{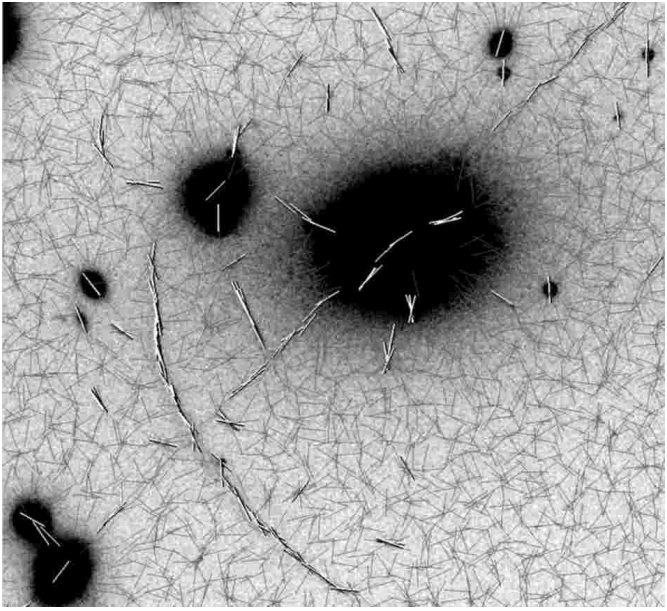}
\caption{Cell orientations are displayed in the same area shown in
  \figurename~\ref{image:a2390_center:paths}. Bold white lines mark
  orientations with $c^i\ge c_\mathrm{th}$.}
\label{image:a2390_center:ellipticities}
\end{figure}

In the following, superscripts will enumerate the cells. In one
dimension, the cells' relative spatial ordering is obviously preserved:
if $x^i_0<x^j_0$, then $x^i_n\le x^j_n$ after any number $n$ of
iterations of the centre computation. This is not generally true in two
or more dimensions, but the number of exceptions remains negligible for
small $n$. Therefore, the set $\mathcal{N}$ of neighbouring cells of a
cell at the end of its path is assumed to be the same as at the
beginning of its path.

Both spatial and directional information should now be used to
determine whether a cell is likely to be part of a valid feature or
not. To this end, we study the correlation of a cell $i$ with its
neighbours $j\in\mathcal{N}$.

The similarity of orientations between cells can be expressed by
\begin{equation}\label{eq:c_d}
  c_d^{ij}=|\vec{e^i}\cdot\vec{e}^j|\;.
\end{equation}
The relative coordinates of the cell $j$ in an orthonormal coordinate
frame centred on cell $i$ with its $x_1$ axis parallel to 
$\vec{e}^i$ are
\begin{equation}\label{eq:delta}
  \vec{\Delta}'=\vec{x}^j-\vec{x}^i\quad\mbox{and}\quad
  \vec{\Delta}=\left(
    \begin{array}{rcr}
       \Delta_1'e_1^i & + & \Delta_2'e_2^i \\
      -\Delta_1'e_2^i & + & \Delta_2'e_1^i \\
    \end{array}
  \right)\;.
\end{equation}
The coordinate $\Delta_2$ measures the distance cell $j$ from a
possible feature through the centre of cell $i$ pointing towards
$\vec{e}^i$. For convenience, we introduce a measure in the
range $[0,1]$,
\begin{equation}\label{eq:c_x}
  c_x^{ij}=
  \begin{cases}
    1-\frac{|\Delta_2|}{d_1} & \mbox{for}\quad|\Delta_2|<d_1\\
    0                        & \mbox{else}\\
  \end{cases}\;,
\end{equation}
where $d_1$ is the initial distance between neighbouring cells.

Using only the orientation and distance correlations $c^{ij}_d$ and
$c^{ij}_x$, it is already possible to estimate the correlation
$c^{ij}=c_d^{ij}c_x^{ij}$ between the cells $i$ and $j$ reasonably well.
Including the entire neighbourhood, this becomes
\begin{equation}\label{eq:c}
  c^i=\frac{1}{|\mathcal{N}|}\,
  \sum_{j\in\mathcal{N}}c^{ij}\;,
\end{equation}
where $|\mathcal{N}|$ denotes the number of cells in the
neighbourhood $\mathcal{N}$.

To reduce the effect of possible errors due to the assumption of an
identical initial and final neighbourhood, and to lower the weight
of closely neighbouring cells, $\mathcal{N}$ can be extended
and another factor $c^{ij}_\Delta$ depending only on $|\Delta|$ can 
be introduced which is small for $|\Delta|=0$, tends to zero for 
$|\Delta|\gg d_1$, and has a maximum in $(0,d_1]$. For a sufficiently
large initial $\mathcal{N}$, the modified correlation measure
\begin{equation}
  c_{\Delta}^i=\frac{\sum_{j\in\mathcal{N}}c_\Delta^{ij}c^{ij}}
                    {\sum_{j\in\mathcal{N}}c_\Delta^{ij}}
\end{equation}
becomes independent of the initial neighbourhood and allows suppressing
the contribution to the correlation of cells very close to cell $i$.
Since it slightly increases the execution time, introduces an additional
degree of freedom and leaves the suitable form for $c^{ij}_\Delta$
to be determined, this distance weighting was ignored here.

Values $c^i$ are computed from (\ref{eq:c}) for each cell, restricting
the neighbourhood to the eight surrounding cells. Cells with
$c^i<c_\mathrm{th}$ are excluded from the further processing. For many
applications, a reasonable value for the threshold is
$c_\mathrm{th}=0.5$. In a next step, those cells above the threshold
which are likely part of a feature are grouped into \emph{objects},
using a very similar method as described above.

\subsection{Object generation}

Each remaining cell $i$ is now individually compared with all other
remaining cells $j$ in its neighbourhood, based on the known
correlation coefficients $c_d^{ij}$ and $c_x^{ij}$. Neighbourhoods are
extended essentially with a slightly modified version of Bresenham's
line-drawing algorithm applied to cells instead of pixels. It generates
an appropriately oriented region, for example with a length of nine and
a width of three cells, surrounding the cell $i$ in its initial
position. To avoid cells not following a possible curvature in the
feature, cells with a large separation $\Delta_1$ can additionally be
handicapped, e.g.~by introducing a factor
\begin{equation}
  c_c^{ij}=0.8+\frac{0.2}
  {1+\frac{1}{200}\left(\frac{\Delta_1}{d_1}\right)^4}\;,
\end{equation}
into the correlation coefficient
\begin{equation}
  C^{ij}=c_d^{ij}\,c_x^{ij}\,c_c^{ij}
\end{equation}
between any two cells $i$ and $j$. If $C^{ij}$ exceeds a threshold
$C_\mathrm{th}$, cell $j$ is added to the object which cell $i$ is
already belonging to. If cell $i$ is not part of an object yet,
an object is created including both cells.

Cells may belong to more than one object, which is reasonable because
objects cannot be properly distinguished before they are completely
defined. When the previous step is completed, any two objects with at
least one common cell are combined into one. Since any averaging between
neighbouring cells was avoided which may not be part of the underlying
feature, the value of $C_\mathrm{th}$ may exceed $c_\mathrm{th}$.
Tests showed that $C_\mathrm{th}=0.7$ is a reasonable choice.

Cells located at the exact same position, e.g.~at the centre of a
radially symmetric feature, will be grouped into one object. Since only
features exceeding a certain length are valid, such objects can later
be removed from any following analysis. Also, objects consisting of
only very few cells are likely to be the result of the random behaviour
of cells far from any feature, and can similarly be invalidated. For
testing, we set the minimum number of cells in an object to four.

\subsection{Object graphs}

The above procedure may result in objects covering several features.
For disconnecting them, the topographical structure of the feature, or
of spatially connected features underlying them, must be taken into
account. Defining objects by concatenation of cells is not practical for
this purpose. Therefore, it is useful to represent their ridge lines by
undirected geometric graphs.

\subsubsection{Initial graph generation}

For each object, graphs are created in the following three steps:

\begin{enumerate}

\item \emph{nodes} are defined from spatially close cells, whose
  averaged position and orientation are assigned to the nodes;

\item if another node is found within a limiting angle of $\pm\pi/4$ of
  a node's direction, it is connected with the node; and

\item separate connected subsets are connected into one connected graph.

\end{enumerate}

For finding the cells to be combined into nodes, each cell is initially
marked as unprocessed. The algorithm loops over all unprocessed cells in
the object and again calculates the absolute cosine between the cells'
directions $c_d^{ij}$, the position $\vec{\Delta}$ of cell $j$ in the
oriented coordinate system of cell $i$, and a measure of the
perpendicular distance
\begin{equation}\label{eq:c2_x}
  {c'}_x^{ij}=
    \begin{cases}
      1-\frac{1}{2}\frac{|\Delta_2|}{d_1} & \mbox{for}\quad 
      |\Delta_2|<2d_1\\
      0                                   & \mbox{else}\\
    \end{cases}
\end{equation}
for every other unprocessed cell $j$ in the object with a spatial 
distance $|\vec{\Delta}|<d_1$, where $d_1$ is again the initial distance
of the cells $i$ and $j$.

Using a third correlation threshold $C'_\mathrm{th}$, the mean positions 
and directional vectors of all cells $j$ with 
$c_g^{ij}=c_d^{ij}{c'}_x^{ij} \ge C'_\mathrm{th}$ are determined, the 
cells $j$ and $i$ are marked as processed, and a new node with these 
properties is created. In ${c'}_x^{ij}$, the perpendicular distance is 
weighted less than in $c_x^{ij}$ before to include cells of high 
orientational correlation in a greater spatial range into one node, 
thus preventing the construction of spatially close parallel nodes 
which might later cause loops in the graph.

The algorithm continues by searching the next unprocessed cell. It is
important to note that the mean of the directional vectors
does not return an average orientation, which must instead be 
calculated as the mean over the normalised ellipticities, e.g.~using
\begin{equation}
  \frac{\vec{\chi}}{|\vec{\chi}|}=(e_1^2-e_2^2,2e_1e_2)\;,
\end{equation}
which can then be used to derive directional vectors. In this way, all
nodes in the graph are created as cell averages. As with
$C_\mathrm{th}$, the threshold $C'_\mathrm{th}$ quantifies the
relation between two instead of more cells, and should thus
exceed $c_\mathrm{th}$. Tests showed
that $C'_\mathrm{th}=C_\mathrm{th}=0.7$ avoids including
substructures but allows features to overlap.

\begin{figure}
  \includegraphics[width=\hsize]{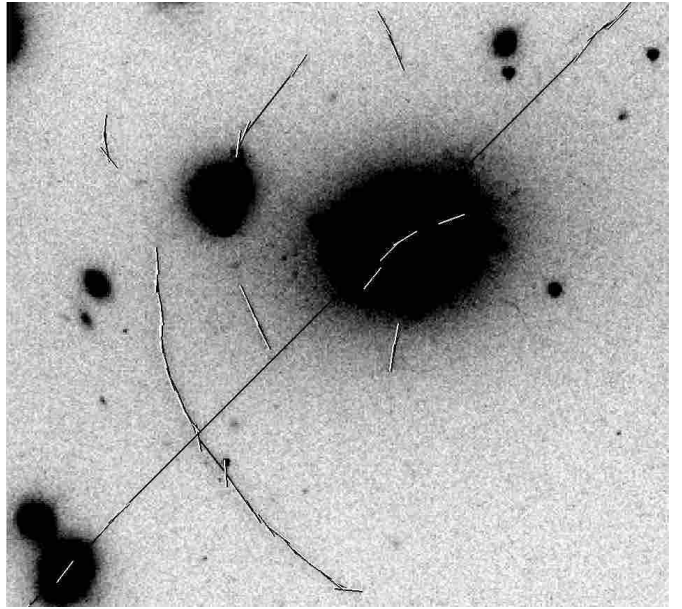}
\caption{Graphs overlaying the final detections. The orientation
  attributed to each node is marked by a black and white line.}
\label{image:a2390_center:graphs}
\end{figure}

To connect the nodes, the relative distance $\vec{\Delta}$ from a node
$i$ to every other node $j$ is calculated using Eq.~(\ref{eq:delta}).
From this, a modified distance
\begin{equation}
  r_e=\sqrt{\Delta_1^2+16\Delta_2^2}
\end{equation}
is computed for each node with
$\left|\frac{\Delta_2}{\Delta_1}\right|$, i.e.~within an angle of
$\pi/4$. The nodes with the lowest $r_e$ among all nodes are connected
to node $i$, if they exist.

The angle $\pi/4$ limits the curvature to less than $\pi/4d_1$. The
metric $r_e$ assigns equal distances from cell $i$ to all points on
an ellipse with an axis ratio of four and a major axis parallel to the
$e_1$ axis centered on $i$. Consequently, nodes with a small
perpendicular distance are connected preferentially. This reduces the
influence due to nodes of features crossing within the same object.

The implementation of this method can lead to a disconnected graph, with
individual features separated by independent structures in the object. A
connected graph can be created by repeatedly connecting the subgraph
containing the first node in the object to the closest subgraph
separated from it. Specifically, an additional variable $v$ is assigned
to each node, which is initially set to zero for all nodes. In a loop,
the value one is recursively propagated to all connected nodes, starting
in node zero. In the next step, the distance $r_e$ is measured between
all nodes $i$ connected with node zero which thus have $v^i=1$, and all
disconnected nodes $j$ with $v^j=0$. The pair of nodes $i$ and $j$ with
the minimal distance is then connected, all $v$ are re-initialised with
zero, and the process is repeated until no disconnected nodes are left.

\subsubsection{Generation of subgraphs from graphs}

The graphs resulting from the preceding procedure are not ideally
suited for post-processing because they may belong to multiple
features, or features contaminated by spurious structures. Therefore,
all their subgraphs which possibly represent either part of or a
complete valid feature are generated using the specific structure of
the graphs representing valid features: each node must be connected to
either one or two other nodes, and the relative angle between two
edges at a node must fall within $\pi\pm\alpha$,
with $0\le\alpha\lesssim\pi/4$ depending on the maximum allowed
curvature. That is, graphs satisfying this definition are either
arc-like, with two end nodes having only one connection and an
arbitrary number of nodes with two edges in between, or ring-like,
having no end nodes and containing only nodes met by two edges.

To find the first variant, all edges connected to a node $i$ are
checked for edges joining on the opposite side of the node within the
specified angle range. If there is no such opposite edge, node $i$ is
a starting or ending point, and the algorithm recursively goes through
all nodes connected through this and the following edges which are
within the required angle interval with their preceding edges, and lead
to nodes not already belonging to the current path. All nodes $j$ in
this recursive scheme for which no successive edge is found within the
angle interval are end nodes. A new graph is created starting with
the path from $i$ to $j$ if an equal graph, consisting either of
the same path from $i$ to $j$ or its exact reverse, does not exist yet.

To find the second variant, one can start the recursion for all edges
in the original graph which are not yet part of any other graph.
The resulting graphs will miss one edge and thus have a starting and
an ending point, which makes further computations slightly easier, but
will equally represent any underlying feature. We tested our algorithm
setting $\alpha=\pi/5$.

\subsubsection{Graph concatenation}

Our algorithm's final step is the concatenation of those new graphs
which may represent a single feature. To this end, a circle is fit to
each graph, and the nearest of any other graphs falling on that circle
is added to it, provided it is closer than three times the angle spanned
by the two lines from the centre of the circle to the first and last
nodes in the original graph.

Then, a new circle is fit to restart the procedure. Since a true
least-squares fit of a circle to a set of points is impossible
analytically, and finding the solution numerically is computationally
expensive, the fit is simplified by assuming that one of the first and
the last two nodes of the graph fall exactly on the circle. The
radius can then be fit analytically, and the best fit in a least-square
sense is chosen from the four solutions. The deviation of this method
from an unconstrained fit is negligible in most cases, although the
method is feasible only because of the already simple form of the
given graphs.

\section{Potential problems}

A number of unresolved problems remain, as well as problems with known
solutions. Both are considered in the following.

\subsection{Ellipticity bias}\label{subsection:ellipticitybias}

Given an odd scale size $d_0$, the straightforward method of measuring
ellipticities for each cell computes the sum
\begin{equation}\label{eq:Qijdiscrete}
  Q_{ij}=\frac
    {\sum_{\vec{x}\in A}(x_i-\bar{x}_i)(x_j-\bar{x}_j)q(I(\vec{x}))}
    {\sum_{\vec{x}\in A}q(I(\vec{x}))}\;,
\end{equation}
inside the rectangle
\begin{equation}\label{eq:Arectangular}
  \begin{aligned}
  &A=\left\{
     \vec{x}\in\mathbb{Z}^2: x^{\rm{min}}_1\le x_1<x^{\rm{max}}_1\wedge
     x^{\rm{min}}_2\le x_2<x^{\rm{max}}_2
     \right\}\\
  &\begin{aligned}
    &\mbox{with}&&
     \vec{x}^{\rm{min}}=\left(
       \bar{x}_1-\frac{d_0-1}{2},
       \bar{x}_2-\frac{d_0-1}{2}
     \right)\\
    &\mbox{and}&&
     \vec{x}^{\rm{max}}=\left(
       x^{\rm{min}}_1+d_0,
       x^{\rm{min}}_2+d_0
     \right)\;,
   \end{aligned}
  \end{aligned}
\end{equation}
where $\vec{\bar{x}}$ are pixel coordinates within the cell. However,
since noise fluctuations in the four corners of $A$ have a stronger
influence on the secondary brightness moments than similar
fluctuations along the edges or in the interior of the cell, the
resulting ellipticities will preferentially be aligned with the
diagonal.

This significant ellipticity bias can be avoided using 
position-dependent weights $q'(I(\vec{x}),\vec{x})=q(I(\vec{x}))\cdot
A_p(\vec{x})$ and modified coordinate components $x'_1(\vec{x})$ and
$x'_2(\vec{x})$ in (\ref{eq:Qijdiscrete}), e.g.
\begin{equation}\label{eq:Qijdiscretewithdisk}
  Q_{ij}=\frac
    {\sum_{\vec{x}\in A}(x'_i-\bar{x}_i)(x'_j-\bar{x}_j)
       q'(I(\vec{x})),\vec{x})}
    {\sum_{\vec{x}\in A}
       q'(I(\vec{x}),\vec{x})}\;,
\end{equation}
where $A_p(\vec{x})$ is the area of overlap between a circle around
$\vec{\bar{x}}$ with radius $\frac{d_0}{2}$ and the horizontally aligned
square of one pixel side length centred on and typically associated with
a single pixel at position $\vec{x}$. The centre of $A_p(\vec x)$ is
$\vec{x'}(\vec{x})$, and the total integration area $A$ is the same as
above. Using only discrete centre coordinates $\vec{\bar{x}}$, these
values can be pre-calculated analytically during the initialisation of
the algorithm (see Appendix \ref{ap:pixeldiskintegration}).

\subsection{Influence of point sources and galaxies}
\label{subsection:pointsourcesandgalaxies}

Several problems are mostly associated with point sources in the
astronomical context, although they can also appear in other
applications.

For a single bright point source, one problem already solved by setting
a minimal object length is the creation of objects from multiple cells
converging to the exact same location. An efficient method to
approximate the length of an object represented by a set of cells is to
first find the cell $j$ farthest from an arbitrary cell $i$, and then
the cell $k$ farthest from cell $j$, using the distance of cell $k$ from
cell $j$ as a length measurement. This is not necessarily the largest
pairwise distance in the object, but it can only be smaller by a factor
of $\sqrt{3}$ (see Appendix \ref{ap:objectlen}).

A closely related problem for very bright point sources is caused by
brightness slopes which may be much more extended than the scale size
$d_0$. Since a fixed number of iterations is used to determine the
nearest local centre-of-brightness, cells do not necessarily complete
the path to this point and measure the secondary brightness moments on
the slope. They will then have orientations parallel to the slope's
gradient and can, depending on the original cell position, generate
spurious detections. This can be avoided by setting limits for the total
spatial displacement of a cell and discarding cells outside from further
computations. Using $q(I)=\mathrm{max}(I-\bar{I},0)$ and assuming a
constant slope, the total spatial displacement is the number of
iterations times $d_0/3$, and less for local brightness maxima with
their negative curvature in the brightness profile. We set the limit to
$8d_0/11$ for testing the algorithm. A welcome side effect of this
approach is the preservation of the valid cells' original
neighbourhoods.

As a consequence of averaging over an area characterised by the scale
size, bright pixels in the vicinity of a valid feature can outshine
it and prevent its detection. Since point-source images must be closer
to a feature than about half the scale size $d_0$ to pull cells away
from it during the centre-of-brightness determination or to influence
the ellipticity measurement, choosing a smaller value for $d_0$ can
avoid this problem at the cost of a lower signal-to-noise ratio.

Clustered point sources can create false detections by imitating
substructures of a valid feature. Since the distinction of
substructure from point-source clustering is impossible after
averaging over $d_0$, which is one of the basic ideas behind the
algorithm, post-processing of the brightness profiles underlying each
object graph is possibly the only way of solving this type of
problem. Even a single point source enhances the likelihood for a
spurious detection by pulling multiple cells towards a single location,
thereby compromising the simple cell-count filter.

Adding a single cell $j$ with sufficient correlations $c^j$ and $C^{ij}$
further away than the minimal object length and in the neighbourhood of
any cell $i$ among the already assembled cells results in a detection.
For farther cells with high spatial correlation, i.e.~placed along the
assembled cells' mean direction, the orientational correlation will
also be increased if they are affected by the point-source image,
causing them to point at the source.

The situation is similar for two point sources, which can create false
detections if their distance in the image exceeds the minimal object
length and if a cell located at the centre of each point source image
can ``see'' the image of the other within its averaging area. Of course,
a larger point-spread function increases the influence of the points
sources.

Searching for arcs, extended diffraction spikes and blooming satisfy
all criteria for valid features and will thus cause spurious detections.
One approach to avoiding them is to determine the location of
diffraction spikes in an independent step by searching for bright point
sources, then masking blooming by its maximal brightness and
approximating the extended point-spread function. If an arc candidate is
found which spatially coincides with a diffraction spike, it can then be
flagged or invalidated.

Searching for prominent arcs, the scale size may be too large for
detecting galaxies, but if the scale size is small enough, they will
also cause spurious detections if they fulfil the criteria for valid
features. They can then be eliminated based on their lower length and
length-to-width ratio of their isophotes. Another possible way to
invalidate them is to compute their radial brightness profiles, which
will generally be shallower for gravitationally lensed objects. If
small arcs are included into the search, however, it may be impossible
to reliably remove them unless observations in multiple frequency
bands allow a colour discrimination.

\subsection{Chip and image boundaries}

When image data from CCDs of different resolution and sensitivity are
combined into one image, e.g.~for the HST/WFPC2 instrument, or if the
image examined is a mosaic, boundaries across the image between
regions of different noise level or mean background intensity may
occur. A noticeable change in the mean background intensity shifts
cells by about $d_0/2$ towards the brighter region, significantly
increasing the cell density in this narrow space and possibly creating
an orientation bias perpendicular to the boundary. Since both
directional correlation and a low spatial distance perpendicular to
each cell's direction are required for the creation of objects, this
case will not necessarily result in spurious detections, although they
will become more likely due to the increase in cell density.

A change in the noise level and similar mean background intensities in
both regions typically only introduces an ellipticity bias in the
direction of the boundary, thereby enhancing the directional
correlation. Using a weight function $q(I)=\mathrm{max}(I-\bar{I},0)$
for the determination of the centre-of-brightness which ignores pixels
with intensities below $\bar{I}$ additionally leads to a cell shift
into the region of higher noise and an increase in cell density. For
these reasons, regions of different noise in a single image may cause
spurious detections. Setting $q(I)=0$ for pixels outside the image can
be seen as a boundary between regions of equal background but with a
noise change equal to the noise in the image. In the current
implementation of our algorithm, false detections at the image
boundaries are avoided by simply invalidating cells straddling
them. Another approach could generate noise outside the image, with
the drawback of decreasing the signal-to-noise ratio for the affected
cells.

\subsection{Noise and background gradient}
\label{subsection:noise}

We measure the sensitivity of the algorithm using a model image of
$1000\times1000$ pixels with intensities drawn from a Poisson
distribution. The signal is a smoothed line segment of a quarter
circle with 500 pixels radius, randomly oriented and shifted by at
most 50 pixels from the image centre. Intensities without noise are
given by $I(\vec{x})=I_0+I_\mathrm{s}\exp\left(-d^2(\vec
x)/2\sigma^2\right)$, where $I_0$ is the background intensity,
$I_\mathrm{s}$ is the peak signal intensity along the ridge of the
line segment, $d(\vec{x})$ is the distance of the pixel at $\vec{x}$
from the line segment and $\sigma$ is a smoothing scale. The actual
intensities in the image are Poisson distributed random numbers with a
mean of $I(\vec{x})$ and standard deviation of $\sqrt{I(\vec{x})}$.

Although the resulting model signal is not identical to a line segment
with intensity $\sqrt{2\pi}\sigma I_\mathrm{s}$ above the background
smoothed with a Gaussian kernel, it is a good approximation, easily
constructed for comparison with other algorithms and equally close to
the typical valid feature in astronomical applications as a line
segment smoothed with a Gaussian.

For each set of parameters, we note both the number of valid
detections $N_\mathrm{valid}$, zero or one, and the number of spurious
detections $N_\mathrm{spurious}$, where a detection was counted as
valid if all nodes in the detected graph were inside a maximum
distance of one $\sigma$ from the arc and the graph was at least 196
pixels long, one fourth of the line segment's length. If more than one
detection fulfilled this criterion, only the first was counted.

For a constant background of $I_0=100$, a central signal intensity of
$I_\mathrm{s}=6$ and $\sigma=10$, the mean number of spurious
detections in ten runs for various scaling sizes is given in 
\tablename~\ref{table:spuriousdetections}, showing the detection of 
short, random structures in the background. For the range of scale 
sizes shown, the number of valid detections was zero.

\begin{table}[!ht]
\caption{Dependence of spurious detections on the scale size.}
\label{table:spuriousdetections}
\centering
\begin{tabular}{l||cccccccc}
\hline
$d_0$          & 5     & 7    & 9   & 11  & 13  & 15  & 17  & 19\\
\hline
$N_{\rm{spr}}$ & 385.5 & 27.9 & 1.1 & 4.8 & 0.5 & 0.3 & 0.1 & 0 \\
\hline
\end{tabular}
\end{table}

\begin{figure}[!ht]
\resizebox{\hsize}{!}{\input{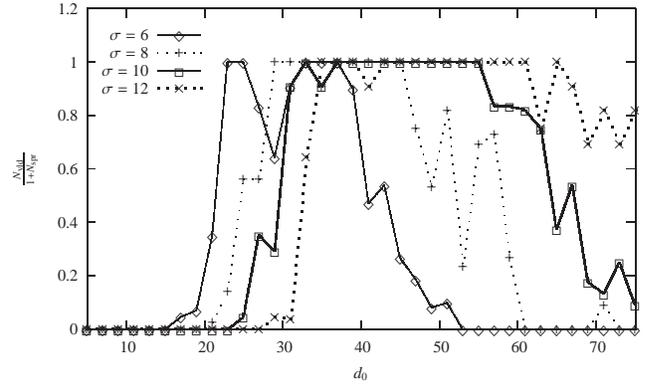}}
\caption{The normalised number of valid detections,
  $N_\mathrm{valid}/(1+N_\mathrm{spurious})$, is displayed as a
  function of the scale size $d_0$ for different smoothing scales
  $\sigma$.}
\label{plot:detections}
\end{figure}

For the same values $I_0$ and $I_\mathrm{s}$, the average detection
ratio $N_\mathrm{valid}/(1+N_\mathrm{spurious})$ found in ten runs is
plotted as a function of the scale size $d_0$ for $\sigma=6,8,10,12$
in \figurename~\ref{plot:detections}. For scale sizes below $2\sigma$, the
cell's areas are dominated by noise even close to the feature,
preventing its detection, and for scale sizes above $10\sigma$, the
feature cannot attract enough cells to sufficiently increase the first
correlation value $c^i$. While there are detections on the line
segment for large scale sizes, these are too short to be valid. With
increasing $\sigma$, the integrated signal intensity increases
proportionally, improving the likelihood for detection, as evidenced
by the larger range of scale sizes with $N_\mathrm{valid}=1$.

For $I_0=100$, $\sigma=10$ and the scale size $d_0=45$ best adapted to this $\sigma$, the average detection ratio $N_\mathrm{valid}/(1+N_\mathrm{spurious})$ in 50 runs for different signal intensities $I_\mathrm{s}$ is given in \tablename~\ref{table:detections}.

\begin{table}[!ht]
\caption{Detection ratio for $I_0=100$, $\sigma=10$ and $d_0=45$
  depending on the central signal intensity $I_\mathrm{s}$.}
\label{table:detections}
\centering
\begin{tabular}{l||cccccccc}
\hline
$I_\mathrm{s}$ &
 4.2 & 4.4  & 4.6  & 4.8  & 5.0  & 5.2  & 5.4  & 5.6\\
\hline
$\frac{N_\mathrm{valid}}{1+N_\mathrm{spurious}}$ &
 0   & 0.03 & 0.12 & 0.38 & 0.42 & 0.63 & 0.89 & 1  \\
\hline
\end{tabular}
\end{table}

In both tests, a noisy but on average constant background was
assumed. If the background has a slope steeper than the intensity
slope up to the ridge line for a background-subtracted feature, cells
cannot find this feature's location during the determination of their
centre-of-brightness, prohibiting detection. In astronomical
observations, bright foreground objects can induce steep intensity
slopes which are generally not well described by a first-order
approximation of the local background, thus offering a challenge to
possible background-subtraction methods.

\section{Implementation specifics}

During the implementation of the above ideas, several arbitrary
choices were made, some of them with a bearing on the results below
in \ref{section:results}, others in an effort to make the code more 
efficient, and some to keep the source code manageable.

\subsection{Initial cell placement}

The most important free parameter, and ideally the only one to be
changed for different image data, is the scale size $d_0$. Based on
it, cells are initially placed on a regular rectangular grid spaced by
$d_1=d_0/2$. Determining the closest centre-of-brightness is done by
applying a discretised version of Eq.~(\ref{eq:xbar}) to find
$\bar{\vec{x}}$ in a rectangular area $A$ as defined in
(\ref{eq:Arectangular}) for each of three iterations. Using a
rectangular area and thereby allowing farther spatial shifts in the
diagonal direction instead of a circular disk for this part of the
algorithm compensates for the initially larger diagonal distance of
the cells.

\subsection{Classes and local indices}

Since the source code was written in C++, it was convenient to use
structures and classes to represent cells, objects and graphs,
providing easy access to data members and hiding most of the necessary
memory management and initialisations from the primary
functions. Since the number of cells remains constant while the
algorithm proceeds, it was sufficient to implement them as
one-dimensional arrays of cell structures, where the original position
and neighbourhood can be readily derived from each cell's index. Graphs
are described by classes for nodes, single graphs, and the list
containing all graphs. To increase the efficiency of the graph
concatenation, whose complexity is $\mathcal{O}(n^2)$ in the number of
graphs $n$, the graph list can be used to subdivide the image into
rectangular regions, each of which maintains a list of all graphs with
nodes in it. These lists can be initialised with linear complexity and
are, combined with a method to mark already processed graphs, used to
enumerate only those graphs in relevant areas. However, while
significantly decreasing the execution time, this does not reduce the
concatenation's squared dependence on the number of graphs.

\subsection{Performance}

Applying the algorithm to a FITS image of $3021\times3021$ pixels with
single (floating point) precision using a scale size $d_0$ of 27
pixels and a very low minimal length of 22 pixels on a personal
computer with one 2.80~GHz processor and 1~GB RAM took approximately
$4.4\,\mathrm{s}$ total execution time of the calculation thread, and
resulted in 137 detected objects prior to applying any later filters.
Apart from setting a minimal object length and cell count, no further
filters are included yet. Most of that time, namely
$\sim3.9\,\mathrm{s}$, was needed for the determination of the
centres-of-brightness and the orientations. The graph concatenation
needed $\sim0.1\,\mathrm{s}$.

Changing the scale size to 13 pixels and reducing the minimal object
length to 10 pixels resulted in 1033 mostly spurious detections of
small features, and increased the execution time to
$\sim11.8\,\mathrm{s}$, of which the determination of the centres of
brightness and orientations needed $\sim5.6\,\mathrm{s}$ and the graph
concatenation another $\sim5.6\,\mathrm{s}$.

Applying the algorithm with the latter parameters to another
single-precision FITS image with $8500\times8300$ pixels and less
small-scale structure gave 395 detections, many of which are
diffraction spikes, and took $\sim161.2\,\mathrm{s}$ in total,
$99.9\,\mathrm{s}$ of which were used for the centre-of-brightness and
orientation computations, and $\sim3.5\,\mathrm{s}$ seconds for the
graph concatenation. The remaining $\sim57.7\,\mathrm{s}$ were used
for transferring data between the threads and
initialisation. Considering that the area is only $\sim8.5$ times
larger, the increase by a factor of $\sim25.6$ in execution time for
the algorithm's first step is mostly caused by extensive memory
swapping, as is the steep increase in the time needed for the data
transfer and initialisations. Apart from conventional optimisation or
using more RAM, this can easily be solved by applying the algorithm
several times on distinct overlapping subregions in the image.

\section{Results}
\label{section:results}

We developed our algorithm primarily using an HST/WFPC2 observation of
Abell~2390 with an exposure time of $2100\,\mathrm{s}$ in the F814W
filter as the base image, which includes multiple arcs. We also tested
the algorithm against other images to confirm its generalisability.

\begin{figure}[!ht]
  \includegraphics[width=\hsize]{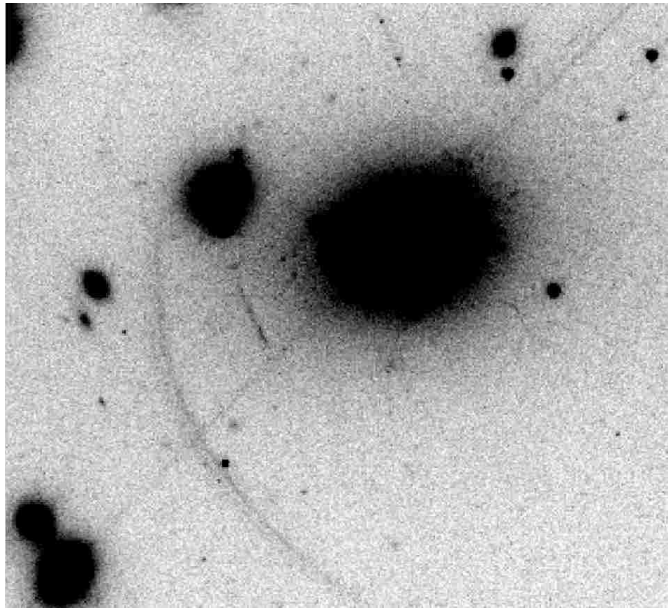}
\caption{This section of the image was used to ``train'' the
  algorithm. It is shown here without additional markers.}
\label{image:a2390_center}
\end{figure}

\begin{figure}[ht!]
  \includegraphics[width=\hsize]{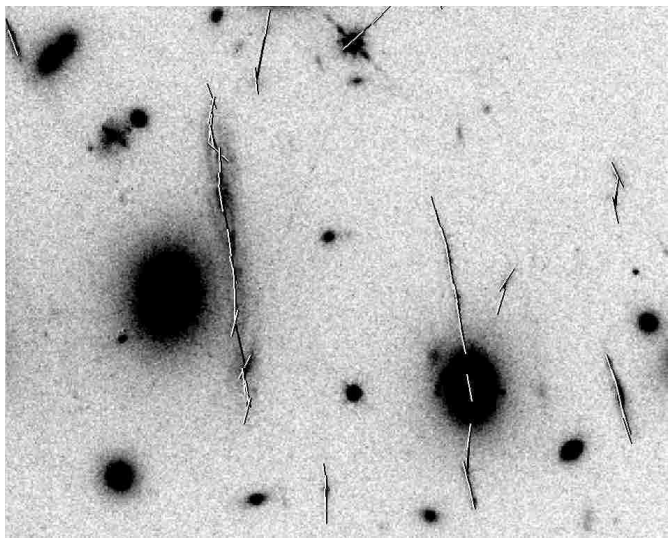}
\caption{The structured arc in the same image is shown with graphs
  overlaying the detections. The scale size is $d_0=27$ here and in
  all of the other HST images shown.}
\label{image:a2390_substructure:graphs}
\end{figure}

\begin{figure}[ht!]
  \includegraphics[width=\hsize]{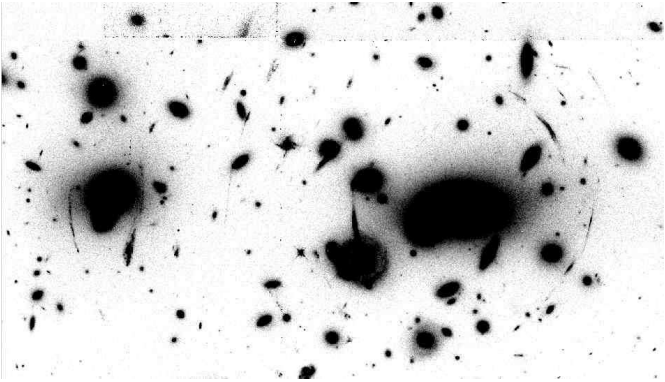}
\caption{Another WFPC2 observation of Abell~2218, taken with a 702~nm
  filter and an exposure time of $1900\,\mathrm{s}$.}
\label{image:a2218}
\end{figure}

\begin{figure}[ht!]
  \includegraphics[width=\hsize]{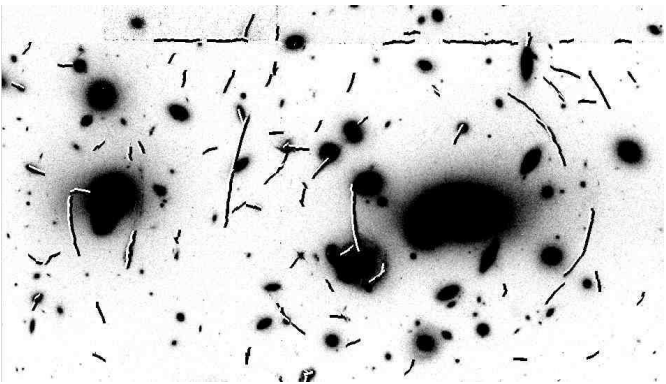}
\caption{This image shows detections marked by black and white lines
  obtained with a scale size of $d_0=25$. Prominent spurious
  detections can be seen near the horizontal border between the WFPC2
  CCDs.}
\label{image:a2218:graphs}
\end{figure}

\section{Future work}

Though it is one aim of the presented algorithm to introduce as few as
possible arbitrary methods and variables, several exist, and they
should be optimised. This refers to the free parameters already
introduced in the text, but also general methods. For example, one
possibility would be to change the initially rectangular grid to a
hexagonal one to provide an equally spaced set of immediate neighbours
and possibly a reduction in necessary overlap.

Point sources can, as detailed in
\ref{subsection:pointsourcesandgalaxies}, significantly increase the
likelihood of spurious detections, which are probably best removed by
a post-processing of those areas in the image underlying detected
graphs. Since these areas represent only a fraction of the complete
image, more elaborate filtering techniques are possible, but as of
now, a filter capable of distinguishing arc substructure from
clustered point sources was not implemented yet. Stellar photometry in
crowded fields already deals with similar issues, and trying to model
a feature as a superposition of point-spread functions, similar to the
approach of the \emph{daophot} program (\cite{dao1987}), could be a
sound if computationally expensive approach. Using the large
elongation of valid features, the width of low-intensity isophotes
could also be used to discriminate features induced by only one or two
point sources, although a slope in the detection's background
increases the minimal isophote intensity, possibly rendering this
method useless and necessitating careful background subtraction.

Diffraction spikes are another problem mentioned in
\ref{subsection:pointsourcesandgalaxies}, and a method to estimate
their positions without specific \emph{a priori} information on the
detector is already implemented, but must be combined with the our
algorithm to remove detections coinciding with them.

We presented a first quantification of the algorithm's sensitivity and
parameter dependence in \ref{subsection:noise}, but the simple way of
modelling a valid feature and the assumption of a constant background
intensity make it difficult or impossible to extrapolate to the
applicability of the algorithm on real datasets from this. Also, the 
observational data presented above was less than comprehensive, since
comparatively high-quality images were used. 
For these reasons, the algorithm will be applied to simulated 
strong-lensing images with varying seeing conditions and detector 
properties as well as to data available from current surveys, 
e.g.~the HST-based \emph{COSMOS} or ground-based observations which, 
given their wider area, are particularly interesting.

As of now, the algorithm uses only images taken in one spectral
filter. However, where sufficiently deep and resolved observations are
available in several spectral bands, this information could be used
both to remove spurious detections separable by the varying spectra of
their components and to increase the algorithm's sensitivity by using
the additional parameter space for correlation measurements.

\section{Summary}

We have presented an algorithm for the fast and automatic detection of
arc-like features in astronomical data. The algorithm proceeds in
three major steps. First, the image pixels are grouped into square
cells which are moved in a fixed number of iteration steps towards their
centre of brightness. They thus find a final position on or near a local
intensity peak. Second, the quadrupole moment of the light distribution
is measured in each cell, where it defines an elongation and a direction.
Third, the neighbourhood of each cell is searched for such cells whose 
intensity distribution points into a similar direction.
If so, the cells are connected to form an object, and the procedure is
continued until all cells have been processed. The objects found in this
way can then be automatically classified.

Having arc-like features in mind which may be just above the noise
limit, we avoid smoothing in the algorithm which may give rise to
subtle biases. Smoothing may make arcs disappear, but it may also
create spurious arc-like features by anisotropic stretching of
positive noise fluctuations. The algorithm also avoids object
detection and uses only the local intensity distribution in the
cells. These design criteria enable an implementation which operates
very fast. For example, it was possible to scan an image of
$3021\times3021$ pixels in less than five seconds on an ordinary
desktop PC.

We used HST images to illustrate the steps of the algorithm and its
efficiency. Remaining potential problems concern the presence of
bright point sources and extended galaxies in the images, boundaries
of images and CCD chips in image mosaics, and residual gradients in
the noise or the image backgrounds, and we outline how they can be
overcome.

Apart from several parameters which control how cells may be connected
with neighbouring cells to form objects, the single main parameter is
the number of pixels to be grouped into a cell. This can be suitably
chosen depending on the resolution and the quality of the image, and
the algorithm is fast enough to allow calibration runs.

The main application which we have in mind for the algorithm is
blindly scanning wide-field images for arcs in order to construct
unbiased arc samples from large data sets. We shall study in a
forthcoming paper how the detection efficiency and reliability will
depend on noise, seeing, and the density of foreground objects.

\begin{acknowledgements}
This work was supported by the Sonderforschungsbereich 439, ``Galaxies
in the young Universe'', of the Deutsche Forschungsgemeinschaft.
\end{acknowledgements}

\appendix

\section{Obtaining an orientation from the complex ellipticity 
$\chi$}\label{ap:orientation}

\begin{figure}
  \includegraphics[width=\hsize]{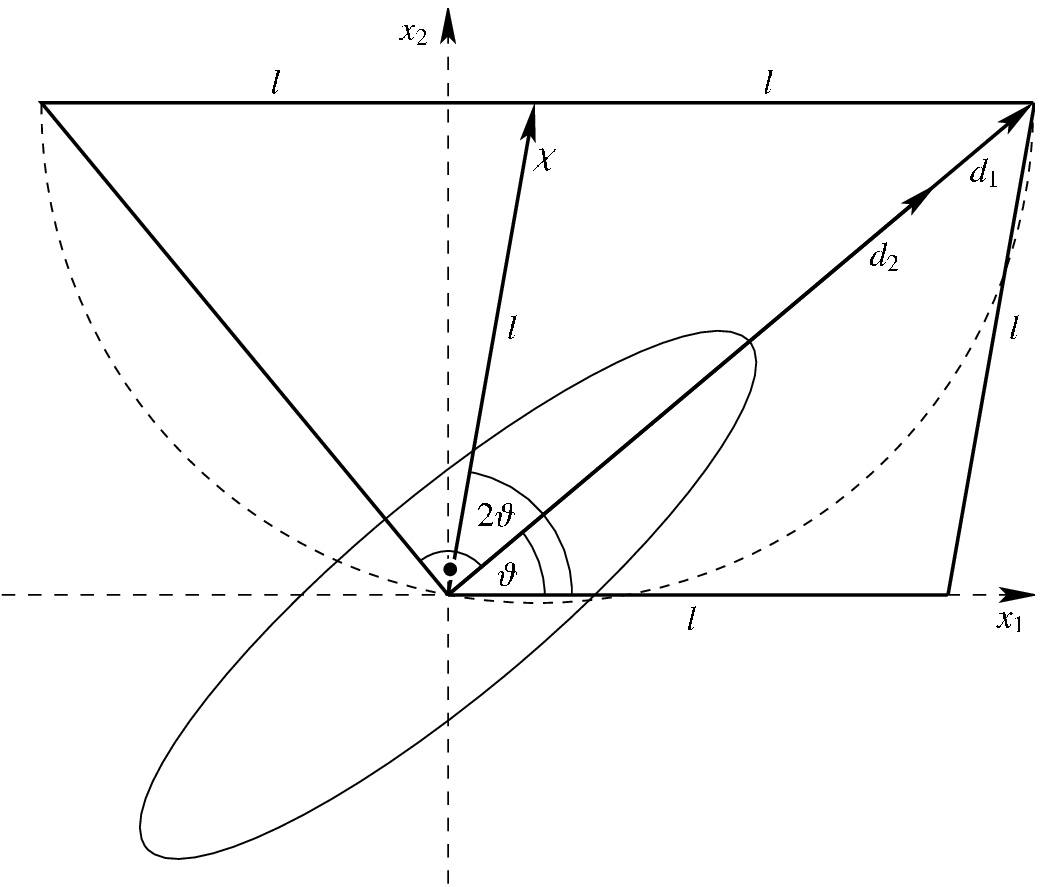}
\caption{An elliptical isophote and the geometric construction of the
  orientation $\vartheta$.}
\label{image:orientation}
\end{figure}

In $\mathds{C}$, the real axis and the complex ellipticity $\chi=
\frac{1-r^2}{1+r^2}\exp(2\ii\vartheta)$ span $2\vartheta$, twice the 
angle $\vartheta$ in image space between the $x_1$-axis and the main 
axis of an underlying ellipse. Writing $(\chi_1,\chi_2)=
(\Re(\chi),\Im(\chi))$, a vector parallel to the ellipse's 
main axis can be obtained as $\vec{d_1}=(\chi_1+|\chi|,\chi_2)$, 
one diagonal of the rhombus with sidelength $l=|\chi|$ seen on the 
right side of \figurename~\ref{image:orientation}. The angle between 
$\vec{d_1}$ and the $x_1$-axis is $\vartheta$. Since $|\vec{d_1}|
\rightarrow0$ for $2\vartheta \rightarrow\pi$, this method cannot be 
used for the bisection of angles $2\vartheta$ close to $\pi$. Of 
course, one can also bisect $2\vartheta$ using $\vec{d_2}=(\chi_2,
|\chi|-\chi_1)$, i.e. the vector $(\chi_1-|\chi|,\chi_2)$ rotated 
clockwise by $\frac{\pi}{2}$, where Thales' theorem using the 
semicircle in \figurename~\ref{image:orientation} can be used to show that 
the angle between $(\chi_1-|\chi|,\chi_2)$ and $\vec{d_1}$ must be 
$\frac{\pi}{2}$. Contrary to $|\vec{d}|$, $|\vec{d_2}|\rightarrow0$ for 
$2\vartheta\rightarrow0$. Using
$$
  \vec{d}=
  \begin{cases}
    (\chi_1+|\chi|,\chi_2) & \mbox{for}\quad\chi_1\ge0 \\
    (\chi_2,|\chi|-\chi_1) & \mbox{for}\quad\chi_1<0   \\
  \end{cases}
$$
as in (\ref{eq:e}), $|\vec{d}|\ge\sqrt{2}|\chi|$ in all cases, making a
reliable computation of the orientation $\vec{e}=
\frac{\vec{d}}{|\vec{d}|}$ possible.

\section{$\sqrt{3}$ relation of the object length and the maximum 
pairwise distance}\label{ap:objectlen}

In \ref{subsection:pointsourcesandgalaxies}, the following method is 
used to determine an approximate object length: First, an arbitrary 
cell $A\in \mathcal{M}$, where $\mathcal{M}$ is the set of all cells 
in the object, is choosen. Then, the cell $B\in \mathcal{M}$ farthest 
from $A$ is found. Last, the cell $C\in \mathcal{M}$ farthest from $B$ 
is determined where $r=|BC|$ is used as the object length. It can be 
shown that $\sqrt{3}r$ is greater or equal to the maximum pairwise 
distance for all cells in the object:

\begin{figure}
  \includegraphics[width=\hsize]{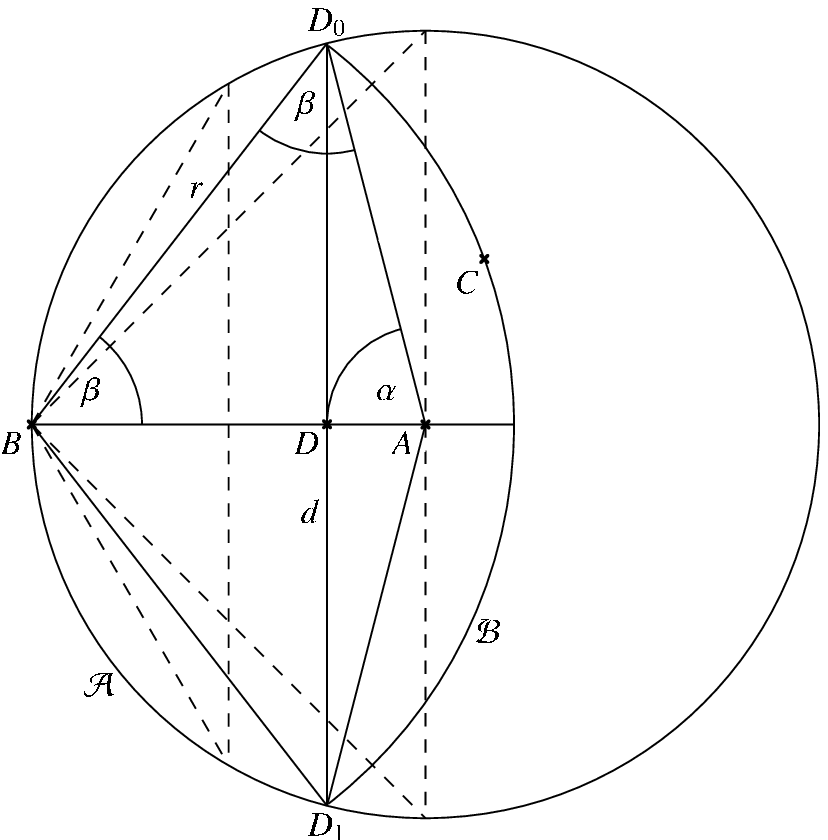}
\caption{The length determination method used in 
\ref{subsection:pointsourcesandgalaxies} requires that the centre 
coordinates of all cells in the object must be in the intersection of 
the circular disks $\mathcal{A}$ and $\mathcal{B}$. It is shown that 
$\sqrt{3}r$ is greater or equal to the maximum pairwise distance of the
cells in the object, which must be lesser or equal then $d$ if $r$ is 
between $|AB|$ (left dashed lines) and $\sqrt{2}|AB|$ (right dashed 
lines). For $r>\sqrt{2}|AB|$ the maximum pairwise distance is limited 
by $2|AB|$.}
\label{image:proofobjectlen}
\end{figure}

Putting the method in more formal terms,
\begin{align}
              & A,B\in \mathcal{M}&&\mbox{with}&&|AB|\ge|AX|\quad
	        \forall X\in \mathcal{M}\quad\mbox{and}\label{eq:AB}\\
              & B,C\in \mathcal{M}&&\mbox{with}&&|BC|\ge|BX|\quad
	        \forall X\in \mathcal{M}               \label{eq:BC}\\
\Rightarrow\; & |BC|\ge|AB|&&\mbox{and}&&|BC|\le |BA|+|AC|\le 2|AB|.
                                                       \label{eq:BCAB}
\end{align}
From \eqref{eq:AB} and \eqref{eq:BC}, all points in $\mathcal{M}$ are 
inside the intersection of the circular disk $\mathcal{A}$ around $A$ 
with radius $|AB|$ and the circular disk $\mathcal{B}$ around $B$ with 
radius $r=|BC|$. The point where the line between the intersections 
$D_0$ and $D_1$ of the boundaries $\partial\mathcal{A}$ and 
$\partial\mathcal{B}$ crosses the straight line $AB$ shall be $D$ and 
the distance between $D_0$ and $D_1$ shall be $d$, as illustrated in 
\figurename~\ref{image:proofobjectlen}. Using \eqref{eq:BCAB}, the problem 
can be separated into three cases:

\begin{enumerate}
\item $r=|AB|$: $\mathcal{A}$ and $\mathcal{B}$ have equal radius, 
 therefore the problem is symmetric and $|BD|=\frac{r}{2}$. From this,
 $(\frac{d}{2})^2=r^2-|BD|^2=r^2-(\frac{r}{2})^2\Rightarrow 
 d=\sqrt{3}r$. Since $\frac{d}{2}>|BD|=|AD|$ and the circle with 
 radius $\frac{d}{2}$ centred on $D$ therefore contains the complete 
 intersection of $\mathcal{A}$ and $\mathcal{B}$, the maximum pairwise 
 distance must be lesser or equal than $d$ and $\sqrt{3}r$ must be 
 greater or equal than the maximum pairwise distance.\\
\item $|AB|<r\le \sqrt{2}|AB|$: $\alpha=\angle{D_0AB}$ and $\beta=
 \angle{ABD_0}=\angle{BD_0A}$ as displayed in 
 \figurename~\ref{image:proofobjectlen}. For $r>|AB|$ follows $\alpha>
 \frac{\pi}{3}$ and therefore $\beta<\frac{\pi}{3}$. Using this, 
 $\cos\beta>\frac{1}{2}$ and $\left(\frac{d}{2}\right)^2=r^2-|BD|^2=
 r^2-r^2\cos^2\beta<\frac{3}{4}r^2\Rightarrow d<\sqrt{3}r$. For example
 using $|BD|=|AB|(1-\cos\alpha)$ and $\frac{d}{2}=|AB|\sin\alpha$ it is
 easily shown that $\frac{d}{2}\ge|BD|>|AD|$ and $d$ must again be 
 greater or equal to the maximum pairwise distance. Consequently, 
 $\sqrt{3}r$ is greater then the maximum pairwise distance.\\
\item $\sqrt{2}|AB|<r\le 2|AB|$: the intersection between the circular
 disks $\mathcal{A}$ and $\mathcal{B}$ is equal to $\mathcal{A}$, 
 therefore the maximum pairwise distance must be lesser or equal to 
 $2|AB|<\sqrt{2}r<\sqrt{3}r$.
\end{enumerate}

\section{Determination of the pixel/disk overlap area $A_p(\vec{x})$ 
and centre $\vec{x'}(\vec{x})$}\label{ap:pixeldiskintegration}

\begin{figure}
  \includegraphics[width=\hsize]{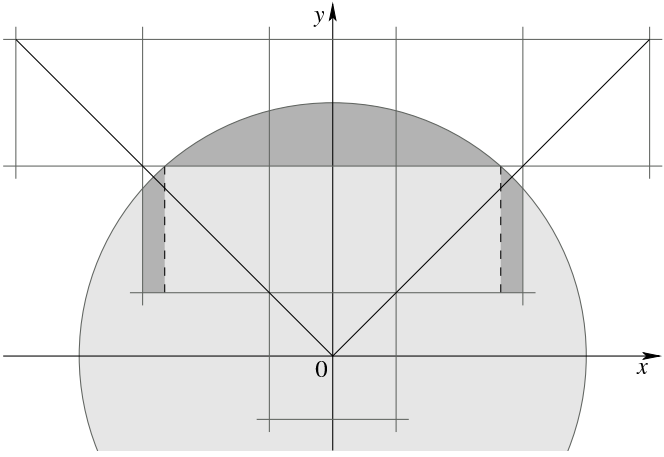}
\caption{Overlap of pixel areas with the area of a circular disk. The
  coordinates $(x,y)$ are chosen so that the disk's centre is at their
  origin and the relevant pixel appears in the upper quadrant enclosed
  by the diagonal lines. To better illustrate the darker areas
  relevant to the integrals (\ref{eq:Awithdisk}) to
  (\ref{eq:Aywithdisk}), an even diameter of four pixels was chosen;
  in the algorithm, an odd diameter equal to the scale size $d_0$ is
  used.}
\label{image:pixeldisk}
\end{figure}

In \ref{subsection:ellipticitybias} the intersection $A_p(\vec{x})$ of 
a circular disk with radius $\frac{d_0}{2}$ around $\vec{\bar{x}}$ and 
the square of one pixel sidelength representing a pixel at position 
$\vec{x}$ must be computed as well as the centre $\vec{x'}(\vec{x})$ 
of this intersection. With a circle radius of $R$, pixel coordinates 
$(x,y)$ chosen such that the circle's centre is at $(0,0)$, $y\ge0$ 
and $|x|\le|y|$, and $A$ being the area inside the pixel square, the 
problem mostly reduces to the following integrals after some 
straightforward case distinctions:
\begin{eqnarray}\label{eq:Awithdisk}
  \iint_A\d x\d y&=&\int_{x_0}^{x_1}\sqrt{R^2-x^2}-y_0\,\d x\\
  &=&\left.\frac{1}{2}\left(
    x\sqrt{R^2-x^2}+R^2\mathrm{arctan}\left(
      \frac{x_1}{\sqrt{R^2-x^2}}
    \right)\right)
  -xy_0\right|_{x_0}^{x_1}\nonumber
\end{eqnarray}
for computing parts of the area $A_p$, and
\begin{eqnarray}\label{eq:Axwithdisk}
  \iint_A\,x\d x\d y&=&\int_{x_0}^{x_1}x\left(
    \sqrt{R^2-x^2}-y_0
  \right)\d x\\&=&\left.
    -\frac{1}{3}(R^2-x^2)^{3/2}-\frac{x^2y_0}{2}
  \right|_{x_0}^{x_1}\nonumber\\
\end{eqnarray}
as well as
\begin{eqnarray}\label{eq:Aywithdisk}
  \iint_A\,y\d x\d y&=&
  \int_{x_0}^{x_1}\int_{y_0}^{\sqrt{R^2-x^2}}y\,\d x\d y\\&=&
  \int_{x_0}^{x_1}\frac{1}{2}(R^2-x^2-y_0^2)\d x=\left.
    \frac{1}{2}x(R^2-y_0^2)-\frac{1}{6}x^3
  \right|_{x_0}^{x_1}\nonumber\\
\end{eqnarray}
for finding the centre $\vec{x}'$. The remaining area integrals have
fixed limits of the form $\iint_A\d x\d y=
\int_{x_0}^{x_1}\int_{y_0}^{y_1}\d x\d y$ and can be included in a
straightforward manner.

\bibliographystyle{aa}
\bibliography{main}

\end{document}